\documentclass[twocolumn,prd,superscriptaddress,showpacs,floatfix, 
preprintnumbers,nofootinbib,showkeys]{revtex4-2}  

\usepackage{amsmath,amsfonts}
\usepackage{epsfig,epsf}
\usepackage{mathrsfs}
\usepackage{graphicx}
\usepackage{latexsym}
\usepackage{amsmath}
\usepackage{amssymb}
\usepackage{slashed}
\usepackage{textcomp}
\usepackage{amsbsy}
\usepackage[colorinlistoftodos]{todonotes}
\usepackage{comment}
\usepackage{ulem}
\usepackage{hyperref}
\usepackage{graphicx}
\usepackage{dcolumn}
\usepackage{caption}
\usepackage{subcaption}
\usepackage{array}
\newcolumntype{P}[1]{>{\centering\arraybackslash}p{#1}}
\usepackage{bm,color}
\usepackage{amssymb}
\usepackage{amsmath}
\usepackage{comment}
\usepackage{makecell,tabularx,multirow}
\usepackage{mathtools}
\usepackage{times}
\usepackage{bigints}
\usepackage{enumitem}
\usepackage{float}
\usepackage{booktabs}
\usepackage{graphicx}
\usepackage{dcolumn}
\usepackage{bm}
\usepackage{comment}
\DeclareGraphicsExtensions{.png,.eps,.pdf}	

\def\beq{\begin{equation}}
\def\eeq{\end{equation}}
\def\barr{\begin{array}}
\def\earr{\end{array}}
\def\dis{\displaystyle}

\begin{document}


\title{Flavour violating charged lepton decays in Little Randall-Sundrum Model }
\author{Akshay A}
 \email{akshay17@iisertvm.ac.in}
\author{Mathew Thomas Arun}%
 \email{mathewthomas@iisertvm.ac.in}
\affiliation{School of Physics, Indian Institute of Science Education and Research, Thiruvananthapuram, India\\
}%

\begin{abstract}
The Little Randall-Sundrum (Little RS) model receives significantly stronger constraints from the flavour observables in comparison to Randall-Sundrum (RS) model. In this paper, we analyse the effect of the electro-weak sector in Little RS on flavour changing decays of charged leptons. We compare the predictions of the model with the current limits on the flavour violating Branching Ratios of $\mu \rightarrow e e e $, $\tau \rightarrow e e e $, $\tau \rightarrow \mu \mu \mu $, $\tau \rightarrow \mu e e $, $\tau \rightarrow e \mu \mu $ , $\mu \rightarrow e \gamma $, and  $\mu Ti \rightarrow e Ti $. And we show that the dominant constraint arises from the $\mu Ti \rightarrow e Ti$ process which strongly limits the KK-1 gauge boson mass ($M_{KK}$) to be $\gtrsim 30.7 TeV $. We then derive and show that generalising the electro-weak gauge sector to include the Brane Localised Kinetic Term (BLKT) relaxes this constraint to $\gtrsim 12 TeV$. Towards the conclusion, we comment on the possibility that the large flavour violating currents can be mitigated by relaxing the assumption regarding the unnatural thinness and rigidity of the UV-brane and discuss the possibility of suppression of these currents in presence of fat fluctuating branes.
\end{abstract}

\maketitle
\section{Introduction}
The charged Lepton Flavour Violation (cLFV) has been in focus ever since  intermediate vector bosons were proposed\cite{PhysRev1101482,Lee:1977tib}. Even with the inclusion of the neutrino oscillation phenomena in Standard Model (SM) of particle physics, the cLFV processes are predicted to be very small. Hence, any evidence of cLFV inevitably point towards New Physics (NP). As of now, these processes are strongly constrained at 90\% C.L. by the Branching Ratios $B_{\mu \rightarrow e \gamma}< 4.2 \times 10^{-13}$\cite{MEG:2016leq}, $B_{\mu \rightarrow e e e}<1.0 \times 10^{-12}$\cite{SINDRUM:1987nra}, $B_{\mu Ti \rightarrow e Ti}< 4.3 \times 10^{-12}$\cite{SINDRUMII:1993gxf}, $B_{\tau \rightarrow e e e}<2.7 \times 10^{-8}$\cite{Hayasaka:2010np}, $B_{\tau \rightarrow e \gamma} < 3.3 \times 10^{-8}$\cite{BaBar:2009hkt}, $B_{\pi^0 \rightarrow \mu e} < 3.6 \times 10^{-10}$\cite{PhysRevLett.85.2877} and $B_{Z \rightarrow \mu e} < 7.5 \times 10^{-7}$\cite{PhysRevD.90.052005}. On the other hand, various beyond SM scenarios do predict such flavour violating processes.

One of the successful extensions of SM in warped extra-dimensions has been the Little Randall-Sundrum (Little RS) model~\cite{Davoudiasl:2008hx,Davoudiasl:2009jk} with a fundamental scale of $\sim 10^3 TeV$.
Recently, the author had studied \cite{DAmbrosio:2020ngh} the correction to $\epsilon_K$ in Little RS arising from the contribution of tree-level KK-1 gluon exchange in $K-\bar{K}$ oscillation. There, it was shown that the strong bound on the imaginary part of the $C_4^{sd}$ operator in the effective Hamiltonian ~\cite{Csaki:2008zd} ruled out the mass scale lower than $\sim 32 TeV$. This constraint was shown to soften significantly in the presence of Brane Localised Kinetic Terms (BLKTs) and Minimal Flavour Protection (MFP)~\cite{Santiago:2008vq} flavour symmetry.

In this paper, we discuss the flavour violation in charged lepton sector of Little RS. We choose to work in a basis that do not mix the left-handed charged leptons. Rather, the entire mixing is taken to be in the right-handed sector, while keeping the 5-D Yukawa coupling anarchic. There can, in principle, exist mixing in the left-handed charged sector through the Tri-bimaximal Cabibo mechanism\cite{King:2012vj}, but such a choice will affect the neutrino mixing matrix as well. For brevity, we do not consider such models here.
Doing this, we can study the flavour violations in the charged lepton sector independent of the neutrino parameters.

We subject the lepton flavour violating decay predictions of anarchic Little RS model to observations from the rare $\mu$ decays such as $\mu \to e \gamma$, $\mu \to 3e$, and $\mu\to e$ conversion in the presence of $Ti$ nuclei and the rare tri-lepton decays of $\tau$.
Among these, we will show that $\mu\to e$ conversion proves to be the most constraining. And, this constraints the lower limit of the KK-1 gauge boson mass to be $\gtrsim 30.7 TeV$. For comparison, the lower mass limit of gauge boson in Randall-Sundrum model~\cite{Randall:1999ee,Randall:1999vf} that satisfy the $\mu \rightarrow e$ conversion is $M_{KK}\gtrsim 5.9 TeV$~\cite{Agashe:2006iy}. 
In line with the discussion on Kaon oscillation in~\cite{DAmbrosio:2020ngh}, here, we study the effect of Brane Localised Kinetic Term (BLKT) in the electro-weak sector of the model, and in particular on the Z boson wave function. We show that this modification relaxes the strong bounds on the KK-1 mass of the gauge boson to $\gtrsim 12 TeV$. Making the KK gauge boson available at the upcoming high energy colliders.
 
For anarchic 5D Yukawa couplings, since the light fermions are localised close to the Ultraviolet(UV)-brane, it is possible that the un-physical assumption of rigid thin branes in Little RS might have a role to play in these large flavour violations. If we replace this thin rigid UV-brane with a fat brane, along with its fluctuations namely branons \cite{Bando:1999di,Hisano:1999bn,Cembranos:2005kv}, then, we show that the flavour violating couplings get suppressed, lowering the bound on KK-1 gauge boson mass to $\gtrsim 10 TeV$.
  
The paper is organised as follows. In the next section, we briefly review the Little RS model, gauge and lepton field KK decomposition and their interactions. In sec.\ref{sec:trileptondecay} we will derive the tri-lepton decay Branching Ratios and $\mu \to e$ conversion rate in the model. Then, we will discuss the effect of BLKT on the interactions and recompute the constraints in sec.\ref{sec:BLKTLRS}. In sec.\ref{sec:summary}, we summarise the work and discuss on the effects of fat branes on the lepton flavour violation.

\section{The Little  RS model}
\label{sec:LRS}
In this section, we briefly recap the Little RS model. Our four-dimensional space-time is assumed to emerge from a 5D AdS space-time, with a fundamental scale $M=10^3 TeV$ \cite{Davoudiasl:2008hx,Dillon:2016fgw}, upon orbifolding on $M_4 \times S^1/Z_2$. The line element of this 5D space-time is given by,
\begin{eqnarray}
ds^{2}= g_{MN}dx^{M} dx^{N} = e^{-2ky}\eta_{\mu \nu}dx^{\mu}dx^{\nu}+dy^{2},
\label{geometry}
\end{eqnarray}
where M,N are 5 dimensional space-time indices, $\eta_{\mu \nu}=diag(-1,+1,+1,+1)$ and $0\le y\le L$. The warp factor is taken to be $k L \sim 7$ so that the warped down scale at the Infrared(IR)-brane becomes $M_5 e^{-k L} \sim {\cal O} (1 \ \text{TeV})$.

The Higgs field, in order to stabilise its vacuum expectation value from quantum fluctuations, is assumed to be localised on this IR-brane. To avoid large localised flavour violating and proton decay operators on the brane, we assume that the gauge fields and fermions propagate in the bulk. Moreover, we consider these fields to transform under the adjoint and fundamental representations of the Standard Model gauge group $SU(3)_C\times SU(2)_{W}\times U(1)_{Y}$ respectively.

Below, we describe the five-dimensional gauge and fermion fields, relevant to the process, and their Kaluza-Klein (KK) decompositions. We begin our discussion with a review of the electroweak sector coupled with the Higgs boson in 5-dimensional space-time.

\subsection{Bulk Gauge Fields}
\label{sec:gauge}
\subsubsection{Action of the 5D Theory}
Let us consider the bulk gauge fields $W_M^a$ and $B_M$, where $M=\{0,1,2,3,5\}$, of $SU(2)_L\times U(1)_Y$, coupled to the scalar localised on the IR-brane. The 5D action for this system is given by,
\beq
   S_{\rm gauge} = \int d^4x\int_{0}^L\!dy\, 
   \Big( {\cal L}_{\rm W,B} + {\cal L}_{\rm Higgs} \Big) \ .
\eeq
The kinetic part of the 5D gauge theory (${\cal L}_{\rm W,B} $) and the Higgs-sector Lagrangian ($  {\cal L}_{\rm Higgs} $) are given by,
\beq\label{Lgauge}
   {\cal L}_{\rm W,B} = \sqrt{g}\,g^{KM} g^{LN}
   \left( - \frac14\,W_{KL}^a W_{MN}^a - \frac14\,B_{KL} B_{MN}
   \right),
\eeq
and 
\begin{eqnarray}
\label{Lhiggs}
   {\cal L}_{\rm Higgs} &=& {\delta(|y|-L)}
   \left[ (D_{M}\Phi)^\dagger\,(D^{M}\Phi) - V(\Phi) \right]\nonumber \\
   V(\Phi) &=& - \mu^2\Phi^\dagger\Phi 
    + \lambda \left( \Phi^\dagger\Phi \right)^2 .
\end{eqnarray} 
The field strength tensor, of the gauge field is denoted by $W_{MN}$, and the covariant derivative by,
\begin{eqnarray}
D_{M}=\partial_{M}-ig_{5}\tau_{a}W_{M}^{a}-ig'_{5}IB_{M} \ ,
\label{covariant}
\end{eqnarray}
where $g_5$ and $g_5'$ are the 5D gauge couplings of W and B bosons, respectively and the Higgs doublet can be decomposed as,
\beq
   \Phi(x) = \frac{1}{\sqrt2}
   \left( \begin{array}{c}
    -i\sqrt2\,\varphi^+(x) \\
    v + h(x) + i\varphi^3(x)
   \end{array} \right) ,
\eeq
with the Higgs vacuum expectation value denoted by $v$, and $\varphi^\pm=(\varphi^1\mp i\varphi^2)/\sqrt2$. 

The $\delta-$function in Eq.\ref{Lhiggs} ensures that the Higgs vacuum expectation value is stabilised to $\sim \mathcal{O}(1 TeV)$. For simplicity, we also perform the usual redefinitions of the gauge fields,
\begin{eqnarray}
W_M^\pm &=& \frac{1}{\sqrt2} \left( W_M^1\mp i W_M^2 \right) , 
    \nonumber\\
Z_M &=& \frac{1}{\sqrt{g_5^2+g_5^{\prime 2}}} 
    \left( g_5 W_M^3 - g_5' B_M \right) ,  \nonumber \\
A_M &=& \frac{1}{\sqrt{g_5^2+g_5^{\prime 2}}} 
    \left( g_5' W_M^3 + g_5 B_M \right) \ .
\label{boson}
\end{eqnarray}

\subsubsection{KK-Decomposition}
After compactification, the Kaluza-Klein decomposition of 5D gauge field becomes,
\beq
\begin{aligned}
   V_\mu(x,y) 
   &= \sum_n V_\mu^{(n)}(x)\,f^{(n)}_V(y) \ , 
\end{aligned}
\eeq
where $V_{\mu}^{(n)}(x) = \{A_{\mu}^{(n)}(x),Z_\mu^{(n)}(x), W^{\pm(n)}_\mu (x) \} $ are the 4-dimensional KK-modes of the photon, Z-boson and the $W^{\pm}$ boson and $f^{(n)}_{V}(y)$ their wave profiles in the bulk. The Euler-Lagrange equation of motion of these bulk modes are given by,
\begin{eqnarray}
-\partial_{5}\left(e^{-2ky}\partial_{5}f^{(n)}_{V}\right)=m^{2}_{n}f^{(n)}_{V} \ .
\label{guageeqn}
\end{eqnarray}
In this equation, we have used $\partial_\mu \partial^\mu V_{\mu}^{(n)}(x) = m_n^2 V_{\mu}^{(n)}(x)$. 
These fields are set to satisfy the boundary condition, $(\delta V^{\mu}\partial_{y}V_{\mu})\big|_{0,L}=0$ and the ortho-normality condition,
\begin{equation}
    \int_{0}^{L}dy f^{(n)}_{V}f^{(m)}_{V}=\delta_{nm}.
\end{equation}
Solving Eq.\ref{guageeqn}, for $m_n=0$, we find that the zero-mode profile of the gauge boson is flat and is given by,
\begin{eqnarray}
f^{(0)}_{V}(y)=\frac{1}{\sqrt{L}}
\end{eqnarray}
For the higher KK-modes ( $m_n \neq 0$), the solution to Eq.\ref{guageeqn} is given in terms of the Bessel $J$ and Bessel $Y$ functions and is of the form,
\begin{equation}
f^{(n)}_{V}(y)=N^{(n)}_{V}e^{ky}\left[J_{1}\left( \frac{m_{n} e^{ky}}{k} \right) +b^{(n)}_{V}Y_{1}\left( \frac{m_{n}e^{ky}}{k} \right)\right],
\label{gaugekkprofile}
\end{equation}
where $N_V^{(n)}$ and $b_V^{(n)}$ are the two constants of integration. While $N_V^{(n)}$ is fixed by the orthonormality condition, $b^{(n)}_V$ is determined using the boundary condition. Demanding Neumann boundary condition at $y=0$, and $y=L$ we get,
\begin{eqnarray}
b_{V}^{(n)} &=& - \frac{J_{0}(\frac{m_{n}}{k})}{Y_{0}(\frac{m_{n}}{k})}  \ \ \text{ at } y=0,\nonumber\\
b_{V}^{(n)} &=& - \frac{J_{0}(\frac{m_{n}}{k}e^{kL})}{Y_{0}(\frac{m_{n}}{k}e^{kL})}  \ \  \text{ at }y=L \ .
\label{bn}
\end{eqnarray}
The mass spectrum of the KK-modes ($m_{n}=x_{n}ke^{-kL}$) can be computed from the solutions to the equation $J_{0}(x_n) Y_0\left(x_ne^{-kL}\right)-Y_{0}(x_n) J_0\left(x_ne^{-kL}\right)=0$, obtained by equating the relations given above. For future convenience we denote the KK-1 gauge boson mass by $M_{KK} = m_1$.

\subsection{Bulk Fermion Fields}
The Clifford algebra in 5-dimensional space-time is defined by gamma matrices, $\Gamma^A = \{ \gamma^0, \gamma^1, \gamma^2, \gamma^3, \gamma^5\}$, that satisfy $\{ \Gamma^A, \Gamma^B \} = 2 \eta^{AB}$, where $\eta^{AB}$ is the flat metric defined on the 5D tangent space. Since the algebra is irreducible, one cannot construct a chirality projection operator in 5 dimensions. Thus the fermion fields constructed in this geometry have 4 complex degrees of freedom, which on compactification leads to vector-like 4-dimensional fermions. Moreover, due to the lack of chiral symmetry, the geometry do not prohibit a mass term in the bulk for the fermions. These mass terms will become crucial for the geometric Froggatt-Nielsen mechanism to generate 4-dimensional fermion mass hierarchy. Let us start our discussion by constructing the 5-dimensional fermionic action.
\subsubsection{5D Fermionic Action}
The 5D action for doublet ($\widehat{\ell}$) and singlet ($\widehat{e}$) leptons can be written as, 
\begin{equation}
S_{\text{fermion}} = S_{\text{kin}} + S_{\text{yuk}}
\label{fermionaction}
\end{equation}
\begin{eqnarray}
S_{\text{kin}}& = &\int d^5 x \sqrt{-g}\, \bar {\widehat{\ell}} \left( \Gamma^A E_A^M D_M + m_{\ell} \right) \widehat{\ell} \nonumber \\ 
&   & + \int d^5 x \sqrt{-g}\, \sum_{e=e,\mu,\tau} \bar {\widehat{e}}  \left( \Gamma^A E_A^M D_M
+ m_{e} \right) \widehat{e} \nonumber  \\
S_{\text{Yuk}}&=& \int d^5 x \sqrt{-g}~  \bar{\widehat{\ell}}_{i}\left( \tilde{Y}_{5D} \right)^{ij}  \widehat{e}_{j}H(x^\mu)\delta(y- L) \nonumber \\ &  &+ h.c. ,
\label{yukaction}
\end{eqnarray}
where $m_{\ell}$ and $m_e$ are the bulk masses for the doublet and singlet lepton fields and $E^{M}_A$ the inverse f\"{u}nfbeins. We denote the tangent space indices with $A,B$ and the 5-dimensional space-time index with $M,N$. Since the f\"{u}nfbeins satisfy the condition, $e^A_M \eta_{AB} e^B_N=g_{MN}$, for the geometry given in Eq.\ref{geometry}, they become $e^{A}_{M}=(e^{-ky}\delta ^{\alpha}_{\mu},1)$.
The five dimensional anarchic Yukawa matrix is denoted as $\Big(\tilde{Y}_{5D}\Big)^{ij}$, with $i,j$ representing the generational indices. In the above equation, $D_M$ represent the covariant derivative in 5-dimensions given by, $D_M =\partial_ M+\omega _M$, where $\omega_{M}=\frac{1}{8} \omega _{MAB}[\Gamma^{A},\Gamma^{B}]$ and the spin connection given by,
\begin{equation}
\omega _{M AB}=g_{RN}E^{N}_{A}\Big(\partial_M E^R_B + \tilde{\Gamma}^R_{MT} E^T_B\Big) \ ,
\end{equation}
where $E^{A}_M$ are the inverse f\"{u}nfbeins and $\tilde{\Gamma}^{R}_{MS}$ are the Christoffel connections. 

Upon compactification, the 5D Dirac fermions decompose to two 4D Weyl spinors. To ensure that only the correct chiral projections survives at the zero-mode, we use boundary conditions,
\begin{equation}
\widehat{\ell}_{L}(++),~\widehat{\ell}_{R}(--),~\widehat{e}_{L}(--),~\widehat{e}_{R}(++) ,
\label{eq:fermionbc}
\end{equation}
at the orbifold fixed points $(y=0, \ y=L)$. Here, $L,R$ stand for the left and right chiral fields under the 4-dimensional chiral projection operator and $+(-)$ stands for the Neumann (Dirichlet) boundary conditions. For example, $\widehat{\ell}_L(++)$ means that we apply the Neumann boundary conditions at both $y=0$ \& $y=L$. 

\subsubsection{The Kaluza Klein decompositions}
 After compactification, the KK expansion of a generic fermion field ($\Psi$) becomes,
\begin{equation}
    \Psi(x,y)_{L,R} = \sum_{n =0}^\infty \frac{1}{\sqrt{L}} \psi^{(n)}_{L,R}(x) f^{(n)}_{L,R}(y,c),
    \label{KKexp}
\end{equation}
where $\psi^{(n)}_{L,R}(x)$ denotes the corresponding four dimensional KK-modes and  $f_{L,R}(y)$ their extra-dimensional profiles in the bulk. These wave profiles are set to satisfy the ortho-normality condition,
\begin{equation}
\int_0^{L} dy \, e^{-3ky} \,f_{L,R}^{(n)}f_{L,R}^{(m)} = \delta_{n,m}.
\label{fermionnorm}
\end{equation}
The normalised zero mode profile for doublets and singlets, with their respective bulk mass parameters $c_{\ell_{i}} = m_{\ell_{i}}/k$ and $c_{e_{i}}=-m_{e_{i}}/k$, can be derived as~\cite{Csaki:2008zd,DAmbrosio:2020ngh},
\begin{eqnarray}
f^{(0)}_{L}(y,c_{\ell_{i}}) = \sqrt{k}f^0(c_{\ell_{i}})\ e^{ky(2- c_{\ell_{i}})}\ e^{(c_{\ell_{i}}- 0.5)k L},\\
f^{(0)}_{R}(y,c_{e_{i}}) = \sqrt{k}f^0(c_{e_{i}})\ e^{ky(2- c_{e_{i}})}\ e^{(c_{e_{i}}- 0.5)k L},
\label{eq:zeromodelepton}
\end{eqnarray}
where,
\begin{equation}
    f^0(c)= \sqrt{\frac{(1-2 c)}{1-e^{-(1-2c)k L}}} \ .
    \label{fermionzeroprofile}
\end{equation}
Using the boundary conditions given in Eq.\ref{eq:fermionbc}, the lightest and next to lightest modes of chiral leptons in , 4 dimensions, become,
\begin{eqnarray}
\Psi_{L}=(\widehat{\ell}_{L}^{i(0)},\widehat{\ell}_{L}^{i(1)},\widehat{e}_{L}^{i(1)}),\nonumber\\
\Psi_{R}=(\widehat{e}_{R}^{i(0)},\widehat{e}_{R}^{i(1)},\widehat{\ell}_{R}^{I(1)}).
\label{KKfstates}
\end{eqnarray}\\


\subsubsection{Yukawa interaction}
\label{subsec:yukawa}
Using the action in Eq.\ref{yukaction} and the wave function for the zero mode leptons given in Eq.\ref{eq:zeromodelepton}, the 4-dimensional Yukawa matrix can be derived in terms of the 5D anarchic Yukawa as,
\begin{eqnarray}
Y_{4D}^{ij}&=& f^{(0)}_{L}(L,c_{\ell_{i}}) Y_{5D}^{ij} f^{(0)}_{R}(L,c_{e_{j}})  \\
&=& \sqrt{\frac{(1-2c_{\ell_i})(1-2c_{e_j})}{(e^{(1-2c_{\ell_i})kL}-1)(e^{(1-2c_{e_j})kL}-1)}} \nonumber\\
&\times& e^{(1-(c_{\ell_i}+c_{e_j}))kL} Y_{5D}^{ij}
\ . \nonumber
\label{4dyuk}
\end{eqnarray}
For the fermion KK-modes given in Eq.\ref{KKfstates}, the above mass matrix becomes,
\begin{equation}
\mathcal{M}=\left(\begin{array}{ccc}
M_0     & M_0F_R    & 0         \\
F_L M_0  & F_L M_0 F_R & M_{KK}    \\
0       & M_{KK}    & 0
\end{array}\right) \ ,
\end{equation}
where the $F_{L,R}^i=\frac{f^{(1)}_{L,R}(L,c_{\ell_i,e_{i}})}{f^{(0)}_{L,R}(L,c_{\ell_i,e_{i}})}$ and $M_0^{ij}=\frac{v}{\sqrt{2}}Y_{4D}^{ij}$. 
Since, the fermions are in the flavour basis, we need to rotate this mass matrix to obtain the physical states. 
In order to do that, it will be easier if we first diagonalise the SM part, $M_0$, with a biunitary transformation $(U_L, U_R)$. Acting on the mass matrix $\mathcal{M}$ with diag($U_L,1,1$) on the left and on the right with diag($U_R^{\dagger},1,1$) we get, 
\begin{equation}
\mathcal{M}=\left(\begin{array}{ccc}
M_D        		    & \frac{v}{\sqrt{2}}\Delta_R  & 0       \\
\frac{v}{\sqrt{2}}\Delta_L & \Delta_1	  		  & M_{KK}  \\
0          		    & M_{KK}     		  & 0
\end{array}\right) \ ,
\label{mass}
\end{equation}
where $M_D = U_L M_0 U_R^\dagger$, $\frac{v}{\sqrt{2}}\Delta_R=U_LM_0F_R=M_DU_RF_R$,
$\frac{v}{\sqrt{2}}\Delta_L=F_LM_0U_R^\dagger=F_LU_L^{\dagger}M_D$ and $\Delta_1=F_LM_0F_R=F_LU_L^{\dagger}M_DU_RF_R$. 

Moreover, since the mixing of higher KK-modes with the zero-mode are suppressed by $\frac{v}{M_{KK}}$, it is convenient and informative to diagonalise the lower $2\times2$ part of the mass matrix. With that, to the leading order expansion in $x=\frac{\Delta_1}{M_{KK}}$, the diagonal mass matrix becomes~\cite{Agashe:2006iy},
\begin{equation}
\mathcal{M}_D=\left(\begin{array}{ccc}
M_D & x_R\frac{1}{2}(1+\frac{x}{4})  & x_R\frac{1}{2}(1-\frac{x}{4})      \\
x_L\frac{1}{2}(1+\frac{x}{4}) & M_{KK}+\frac{\Delta_1}{2}  		  & 0 \\
x_L\frac{1}{2}(1-\frac{x}{4})        		    & 0 		  & -M_{KK}+\frac{\Delta_1}{2}
\end{array}\right) \ ,
\label{eq:diagonalfmasses}
\end{equation}
where $x_{L,R}=\frac{v}{\sqrt{2}}\Delta_{L,R}$. Note that the off diagonal elements in this matrix are very small compared to $M_{KK}$. 
Now, the degeneracy in the KK-1 mode is lifted to,
\begin{eqnarray}
M_{KK}^{(1)}=M_{KK}+\frac{\Delta_1}{2} \ , \nonumber \\ 
M_{KK}^{(2)}=-M_{KK}+\frac{\Delta_1}{2} \ .
\label{kkmixf}
\end{eqnarray}
 
\subsubsection{Couplings with Z boson}
Before discussing the flavour violating effects, we need to identify the relevant couplings of the gauge boson. For simplicity, we will consider the couplings of fermion bilinear with an abelian gauge field in the bulk of AdS. Generalisation to flavour violating interactions of the non-Abelian gauge field is then straightforward. A 5-dimensional action for the $U(1)$ gauge field can be written as,
\begin{eqnarray}
\mathcal{S}=- \frac{1}{4g_{5}^{2}} \int d^{5}x\sqrt{-g}\left(g^{CM}g^{DN}F_{CD}F_{MN} \right) \ ,
\label{Z boson}
\end{eqnarray}
where $F_{MN}=\partial_{M}A_{N}-\partial_{N}A_{M}$ is the field strength tensor and  $g_{5}$ the 5D gauge coupling.\\
In the unitary gauge ($A_5=0$), the vector field can be Fourier expanded as, 
\begin{equation}
    A_{\mu}(x,y)= \sum_{n}A_{\mu}^{(n)}(x) f^{(n)}_{A}(y) \ ,
\end{equation}
and the coupling of the zero-mode lepton bi-linear with the gauge KK-modes can be computed from the overlap integral, 
\begin{equation}
g_{L,R}^{(n)}(c_{\ell ,e})=\frac{g_5}{L} \int_0^{L} dy \,e^{-3 k y} f_A^{(n)}(y) f_{L,R}^{(0)}(y,c_{\ell,e}) f_{L,R}^{(0)}(y,c_{\ell,e}) 
\label{GaugeFOFO}
\end{equation}
In the above equation, $f_A^{(n)}$ and $f_{L,R}^{(0)}$ are as given in Eq.\ref{gaugekkprofile} and Eq.\ref{fermionzeroprofile} respectively.
Since the geometric Froggatt-Nielsen mechanism requires distinct bulk mass values `$c_{\ell,e}$' for the leptons, the couplings $g^{(n)}_{L,R}$ are different depending on the localisation of the fermion zero mode. Replacing the abelian gauge field with the Z-boson, we can explicitly write these interactions as,
\begin{equation}
\mathcal{L}=g^{(n)}_{L}(c_{\ell_i})\bar{\widehat{\ell}}_{L}^{i(0)}\gamma_{\mu}Z^{\mu(n)}{\widehat{\ell}}_{L}^{i(0)}
+g^{(n)}_{R}(c_{e_i})\bar{\widehat{e}}_{R}^{i(0)}\gamma_{\mu}Z^{\mu(n)}{\widehat{e}}_{R}^{i(0)} \ .
\label{coupling}
\end{equation}
These couplings, being depended on the bulk mass parameter `$c_{\ell,e}$', generate flavour violations in the interactions of the gauge boson KK-modes on rotating the fermions to their mass basis. With this understanding, we can now address the consequences of such terms in Little RS.


\section{Flavour violating decays in Little RS}
\label{sec:trileptondecay}

\begin{figure}[h!]
\centering
\vspace{-1cm}
    \includegraphics[width=8cm,height=8cm]{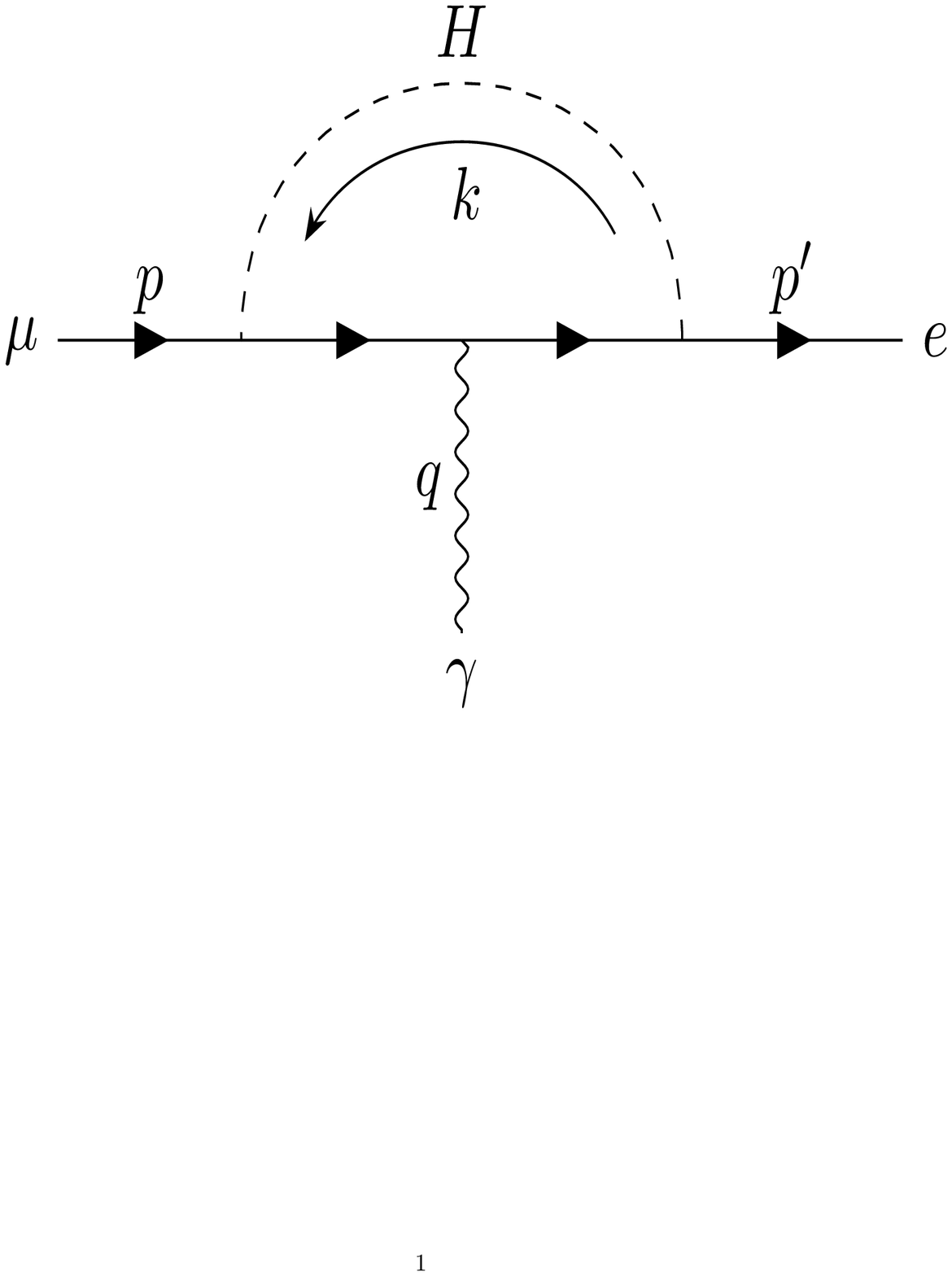}
    \vspace{-3cm}
\caption{The diagram that generates the process $\mu \to e \gamma$.}
\label{fig:meg}
\end{figure}

In this section, we focus on the charged lepton flavour violating decay processes such as  $\mu \rightarrow e \gamma$, $\mu^- \rightarrow e^+e^-e^-$, $\mu \ Ti \to e \ Ti $, $\tau^- \to e^- e^- e^+$, $\tau^- \to \mu^- e^-e^+$ and $\tau^- \to e^- \mu^- \mu^+$, in the Little RS framework.
Among these, the only loop process is the $\mu \to e \gamma$. Though this decay, is divergent in Randall-Sundrum model with brane localised Higgs, since Little RS is an effective theory with much lower cut off $\sim 10^3 TeV$, we expect a need to re-analyse this decay.
The dominant contribution to the process $\mu \rightarrow e \gamma$ proceeds through a 1-loop Feynman diagram with brane localised Higgs and KK-fermions fields as shown in Fig.\ref{fig:meg}. 
The amplitude of the process, assuming $M_{KK}^{(i)}$ to be much greater than the energy scales involved, thus becomes,
\begin{eqnarray}
   A_{\mu \rightarrow e \gamma} &=& \dis \bar{u}(p^{\prime})\{e A_\mu \sum_{i} \frac{1}{M_{KK}^{(i)4}} Y_{e i}  \nonumber \\
   &\times & \dis \int \frac{d^4 k}{(2\pi)^4} \frac{(\slashed{p}^{\prime}+\slashed{k}+M_{KK}^{(i)})\gamma^\mu (\slashed{p}^{\prime}+\slashed{k}+M_{KK}^{(i)})}{k^2-m_H^2} \nonumber \\
   &\times& \dis Y_{i \mu}\} u(p) \nonumber \\
   &=& \dis \frac{1}{2 m_{\mu}} \bar{u_L}(p^{\prime}) \sigma^{\mu \nu} F_{\mu \nu} u_R(p) C_L(q^{2}) \nonumber \\
   &+& \dis \frac{1}{2 m_{\mu}} \bar{u_R}(p^{\prime}) \sigma^{\mu \nu} F_{\mu \nu} u_L(p) C_R(q^{2})\ .
\label{Amutoegamma}
\end{eqnarray}
The divergence in this amplitude come through the large number of fermion KK-modes that contribute to this loop $\frac{C_{L,R}}{m_{\mu}^2} \sim \frac{1}{16 \pi^2} log(N_{KK})$. Their contributions worsen at two-loops. Thus, the cut-off dependent part of the Wilson Coefficient has the form~\cite{Agashe:2006iy}, 
\begin{eqnarray}
\frac{C_{L,R}}{m_\mu^2} &\sim& \dis  \frac{1}{16\pi^2}\left(\frac{Y_{5D}}{M_{KK}}\right)^2\left\{log\left(\frac{\Lambda_{5D}}{k}\right) \right. \nonumber \\ 
&+&  \dis \left.\frac{1}{16\pi^2}Y_{5D}\frac{\Lambda_{5D}^2}{k^2}+....\right\} \ ,
\label{eq:cutoffwilsoncoeff}
\end{eqnarray}
where the first and second terms are the 1-loop and 2-loop contributions respectively. 

Randall-Sundrum model, being UV complete, requires $\Lambda_{5D} = M_{pl} = 10^{16} TeV$ to avoid hierarchy.  Moreover, to avoid quantum gravity effects the curvature should satisfy the condition $k/M_{pl} \ll 0.1$. Hence, assuming $k \lesssim 10^{15} TeV$, the number of KK modes that contribute to the process becomes $N_{KK} \gtrsim 10$. At 2-loop, dimension analysis suggests that the amplitude becomes $\sim N_{KK}^2 = (\frac{\Lambda_{5D}}{k})^2 \gtrsim 100$~\cite{Agashe:2006iy}. Which means that the 1-loop and 2-loop contributions in Eq.\ref{eq:cutoffwilsoncoeff} are of the same order ($Log(\frac{\Lambda_{5D}}{k}) \sim \frac{Y_{5D}}{16 \pi^2} \frac{\Lambda_{5D}^2}{k^2} \sim 1$). And the higher loops contribute strongly and the result is not convergent. This feature threatens any reliable calculations of the $\mu \to e \gamma$ process in Randall-Sundrum model. 

In comparison, Little Randall-Sundrum model is an effective theory with a scale $10^3 TeV$, such that quantum gravity effects are insignificant. It is straightforward to find a parameter space, for example $\Lambda_{5D} \sim 10^3 TeV$ and $k \sim 800 TeV$, in which the number of KK-fermions contributing to the loop can be much smaller ($N_{KK} \sim \frac{1000}{800} \sim \mathcal{O}(1)$), and the cut-off dependent part of the Wilson Coefficient becomes,
\begin{eqnarray}
\frac{C_{L,R}}{m_\mu^2} &\sim& \dis \frac{1}{16\pi^2}\left(\frac{Y_{5D}}{M_{KK}}\right)^2\left\{0.2+ 0.008+....\right\} \ ,
\end{eqnarray}
where the 2-loop term is much smaller than the 1-loop with the higher loop terms further suppressed. Though the amplitude is calculable now, still, the result is sensitive to the cut-off scale of the model, but this can be cured if we can dynamically stabilize this scale. Being phenomenological in nature, Little RS requires new physics to do this. One way this could be achieved is by embedding the model in a six-dimensional $S^1/Z_2 \times S^1/Z_2$ scenario, for which all the radii are stabilised~\cite{Arun:2015kva,Arun:2016csq}.

The cut-off independent part of the Wilson Coefficient, $C_{L,R}(q^2=0)$, can be derived from Eq.\ref{Amutoegamma} as,
\begin{equation}
C_{L,R}(q^{2}=0)=\frac{e m_{\mu}}{32 \pi^{2}} \sum_{i} Y_{e i} \frac{m_{H}^{2}}{ M_{K K}^{(i)3}} Y_{i \mu} \ ,
\label{wilson}
\end{equation}
where $Y = \mathcal{M_D}/(v/\sqrt{2})$ is the rotated Yukawa coupling and $M_{KK}^{(i)}$ are the masses of the first KK-mode given in Eq.\ref{eq:diagonalfmasses} and Eq.\ref{kkmixf} respectively. The branching ratio for this process now becomes~\cite{Chang:2005ag,Moreau:2006np},
\begin{equation}
B(\mu \rightarrow e \gamma) = \frac{12 \pi^2}{(G_F m_\mu^2)^2} \Big[|C_L(0)|^2+|C_R(0)|^2 \Big] \ ,
\end{equation}
where, $C_L(0)$, and $C_R(0)$ can be derived by using Eq.\ref{eq:diagonalfmasses} and Eq.\ref{kkmixf} in Eq.\ref{wilson},
\begin{eqnarray}
C_L(0) &=& \dis e\frac{m_\mu m_H^2}{32 \pi^2} [Y_{e1}Y_{1\mu} \frac{1}{M_{KK}^{(1)3}}+ Y_{e2}Y_{2\mu} \frac{1}{M_{KK}^{(2)3}}] \nonumber \\
&=& \dis e\frac{m_\mu m_H^2}{32 \pi^2 M_{KK}^4} [ \Delta_R \Delta_1 \Delta_L]_{e\mu} \ ,  \nonumber \\
C_R(0) &=& \dis e\frac{m_\mu m_H^2}{32 \pi^2} [Y_{e1}Y_{1\mu} \frac{1}{M_{KK}^{(1)3}}+ Y_{e2}Y_{2\mu} \frac{1}{M_{KK}^{(2)3}}]^\dagger \nonumber \\
&=& \dis e\frac{m_\mu m_H^2}{32 \pi^2 M_{KK}^4} [ \Delta_R \Delta_1 \Delta_L]^\dagger_{e\mu} \ .
\label{wilsoncontribution}
\end{eqnarray}
$\Delta_L$ and $\Delta_R$ are the off-diagonal terms in the fermion mass matrix 
and $\Delta_1$ is the Yukawa mass of the KK-1 leptons given in Eq.\ref{mass}.

This finite part can be computed and on comparing it with the experimental bound on the Branching ratio $B_{expt}(\mu \rightarrow e \gamma) \lesssim 0.042 \times 10^{-11}$~\cite{Workman:2022ynf}, we obtain the lower limit on the mass scale $M_{KK} \gtrsim 1.4 TeV $. 


\subsection{Tri-lepton decays and $\mu-e$ conversions}
Before we discuss the tri-lepton decays in Little RS model, to study the flavour violating effects, it is important and insightful if we identify the model independent four-fermion interactions that contribute to the processes. Adopting the parametrisation in \cite{Kuno:1999jp,Chang:2005ag}, the most general low-energy effective, dimension-6, Lagrangian responsible for these processes can be written as,
\begin{equation}
\begin{split}
-\mathcal{L_{{\rm eff}}}=&\frac{4G_F}{\sqrt{2}}\left[g_3^{ij}(\bar{e}_{iR}\gamma^\mu e_{jR})(\bar{e}_{
kR}\gamma_\mu e_{kR}) \right. \\ & + \left.g_4^{ij}(\bar{\ell}_{iL}\gamma^\mu \ell_{jL})(\bar{\ell}_{kL}\gamma_\mu \ell_{kL})\right. \\ & + \left.g_5^{ij}(\bar{e}_{iR}\gamma^\mu e_{jR})(\bar{\ell}_{kL}\gamma_\mu \ell_{kL}) \right. \\ &+ \left.g_{6}^{ij}(\bar{\ell}_{iL}\gamma ^\mu \ell_{jL})(\bar{e}_{kR}\gamma_\mu e_{kR})\right] + {\rm h.c.} \ ,
\end{split}
\label{ngeq}
\end{equation}
where $g_{3,4,5,6}$ are dimensionless Wilson coefficients.
Note that, we have only considered vector operators and not scalar or pseudoscalar ones. This is because the Higgs contribution to the flavour violating process is suppressed by small masses of the fermions involved. Moreover, the next to leading order effects from KK-fermions mixing are further suppressed.

Like fermions, the electro-weak symmetry breaking with brane localised Higgs boson also mix the KK-levels of the Z-boson. Details of the symmetry breaking and the mass matrices of electro-weak gauge bosons are given in in  Appendix \ref{sec:gaugemass}. To diagonalise this mass matrix, we need to rotate the gauge basis $(Z^{(0)},Z^{(1)})$ to the physical basis $(Z_{(0)},Z_{(1)})$, wherein the admixture enters as,
\begin{equation}
Z_{(0)}=Z^{(0)}+f\frac{m_Z^2}{M_{KK}^{2}}Z^{(1)}, \quad Z_{(1)}=Z^{(1)}-f\frac{m_Z^2}{M_{KK}^{2}}Z^{(0)} \ ,
\label{gaugemix}
\end{equation}
where $f~(\sim\sqrt{2kL})$ parameterises the mixing between the zero and first KK level. From Eq.\ref{GaugeFOFO} it is clear that, since the extra-dimensional wave profile of $Z^{(0)}$ is flat, it couples democratically to all the lepton generations. Whereas, couplings of the fermion zero-mode bilinear with the $Z^{(1)}$ is determined by the appropriate overlap integral and is dependent on the bulk mass parameter $c_{\ell_i,e_i}$,
\begin{equation}
\alpha_i = 2\sqrt{2\pi}\int_{0}^{L} dy \, e^{-3ky}   f^{(1)}_Z(y) [f^{(0)}_{L,R}(y, c_{\ell_i,e_i})]^2,
\label{alphadef}
\end{equation}
In the above equation, $\alpha_e$, $\alpha_{\mu}$, and $\alpha_{\tau}$ denote the ratios of couplings given in Eq.\ref{coupling} to the SM ones. On rotating to the mass basis of leptons, 
the matrix which describes the $Z^{(1)}_{\mu}$ couplings take the form,
\begin{equation}
\mathcal{L}_{int}= g_{L,R} \bar{\Psi} U^{\dagger}_{L,R} \alpha  U_{L,R}  \gamma^\mu
 \Psi Z^{(1)}_{\mu} \ ,
\end{equation}
where $g_{L,R}$ are the SM gauge couplings, $U_{L,R}$ are the unitary mixing matrices for charged SM leptons, $\alpha ={\rm diag}(\alpha_e,\alpha_{\mu},\alpha_{\tau})$ and 
\begin{equation}
 \Psi=\left(\begin{array}{l} e_{L,R} \\ \mu_{L,R} \\ \tau_{L,R} \end{array}\right)
\end{equation}
It can easily be seen that the generational dependence of the $Z^{(1)}$ boson couplings, in Eq.\ref{alphadef}, conspire to create off-diagonal elements that generate flavour violations. Moreover, this property of the KK-1 mode of the Z-boson is inherited by the physical $Z_{(0)}$ boson due to the admixture. Thus the dominant flavour violation in the electro-weak sector is mediated by the zero-mode of the physical Z-boson and integrating them out from low-energy processes results in the Wilson Coefficients shown in Eq.\ref{ngeq}. A detailed discussion is given in Appendix\ref{subsec:coupbr}.

Now, using the operators in Eq.\ref{ngeq} in Eq.\ref{bfracs1}, the branching ratio for the process $\mu \to 3e$ can be written as,
\begin{equation}
BR(\mu \rightarrow 3e) = 2\left(|g^{\mu e}_3|^2+|g^{\mu e}_4|^2\right)+|g^{\mu e}_5|^2+|g^{\mu e}_6|^2 \ ,
\label{bfracs}
\end{equation}
where we have assumed $BR(\mu \to e\nu\nu)=1$.

And, the $\mu-e$ conversion rate~\cite{Kuno:1999jp} becomes,
\begin{equation}
B_{conv} = \frac{2p_e E_e G_F^2 m_{\mu}^3 \alpha_{QED}^3 Z_{eff}^4 Q_N^2}{\pi^2 Z \Gamma_{capt}} \left[|g_{R}^{\mu e}|^2
  +|g_{L}^{\mu e}|^2\right] \ ,
\label{ueconv}
\end{equation}
where the couplings $ g_{3,4,5,6}$, $g_{L,R}$ are given in Appendix(\ref{subsec:coupbr}), $\alpha_{QED}$ is the QED coupling strength and the remaining atomic physics 
constants are given in~\cite{Kuno:1999jp}.

In order to constraint the Little RS model, we impose the following current PDG limits: $BR(\mu \rightarrow 3e) < 10^{-12}$~\cite{SINDRUM:1987nra,Zyla:2020zbs}; $B_{\mu Ti\to e Ti} < 4.3 \times 10^{-12}$~\cite{SINDRUMII:1993gxf,Zyla:2020zbs}.  And for the rare tau decays, we employ the constraints $BR(\tau \rightarrow e^+e^-e^+) < 2.7 \times 10^{-8}$, $BR(\tau \rightarrow \mu^+\mu^-\mu^+) < 2.1 \times 10^{-8}$, $BR(\tau \rightarrow \mu^-e^-e^+) < 1.8 \times 10^{-8}$~\cite{Zyla:2020zbs}. For computing the $\mu \to e$ conversion, we use the numerical values for Titanium from~\cite{wintz1998prepared}. 

Comparing these limits with Eq.\ref{bfracs} and Eq.\ref{ueconv}, we can derive the lower bound on the mass scale in the model.
We present the constraints on $M_{KK}$ for Little RS and compare it with the bounds obtained in Randall-Sundrum model in Table~\ref{table1}. The details regarding our numerical analysis is given in Appendix\ref{subsec:appendix}.
\begin{table}[htbp]
\renewcommand{\arraystretch}{1}
	\begin{tabular}{|m{2cm}|m{2cm}|m{2cm}|m{2cm}|}
		    \hline
		    & & & \\[-1ex]
Model&$BR(\mu\to3e)$ &$B_{conv}$&$BR(\mu\to e\gamma)$\\
& & & \\[-1ex]
\hline
& & & \\[-1ex]
RS&$2.5TeV$&$5.9TeV$&$^*8TeV$\\[1ex]
\hline
& & & \\[-1ex]
Little RS&$20.8TeV$&$30.7TeV$ &$1.4 TeV$ \\[1ex]
\hline
\end{tabular}\\[2ex]
\begin{tabular}{|m{2cm}|m{2cm}|m{2cm}|m{2cm}|}
		    \hline
		    & & & \\[-1ex]
Model&$BR(\tau\to3e)$& $BR(\tau\to3\mu)$&$BR(\tau\to\mu ee)$\\ 
& & & \\[-1ex]
\hline
& & & \\[-1ex]
RS&$0.1TeV$ &$0.4TeV$&$0.36TeV$ \\[1ex]
\hline
& & & \\[-1ex]
Little RS&$2.48TeV$ &$2.43TeV $ &$ 2.50 TeV$ \\[1ex]
\hline
\end{tabular}
\caption{Constraints on the first KK-mode mass, $M_{KK}$, coming from various measurements for a brane Higgs field in both the RS and the Little RS. Except for the $BR(\mu\to e\gamma)$ in the RS model where we have used bulk Higgs, since the brane Higgs is not computable in the RS. }
\label{table1}
\end{table}

\section{Brane Localised Gauge Kinetic Terms}
\label{sec:BLKTLRS}
Going beyond the simplest possible extension of SM in Little RS, the action of the gauge field in 5-dimensions, given in Eq.\ref{Z boson}, can be generalised ~\cite{Georgi:2000ks, Carena:2002me, DAmbrosio:2020ngh,Fichet:2019owx} as, 
\begin{equation}
\begin{split}
\mathcal{S}=&-\frac{1}{4g_5^2}  \int d^{5}x\sqrt{-g}\left\{\left( g^{AM}g^{BN}F_{AB}F_{MN} \right)\right.\\  
&+ \left.\left(l_{UV} \delta(y)+ l_{IR} \delta(y-L)\right)g^{\alpha \mu }g^{\beta \nu}F_{\alpha \beta}F_{\mu \nu}\right\}
\label{ZAction}
\end{split}
\end{equation}
where $l_{UV}$ and $l_{IR}$ are the localised kinetic term strengths at the UV and the IR branes respectively. The origin of these terms are for the time being unknown, but it is understood that for correct re-normalisation of the model such terms are necessary \cite{Georgi:2000wb,Arkani-Hamed:2001uol}. These terms can be hypothesised to have their origin in the unperturbative effects of brane localised matter coupled to the gauge field.

Varying the action in Eq.\ref{ZAction} with respect to the field, the equation of motion becomes,
\begin{equation}
-\partial_{5}(e^{-2ky}\partial_{5}f^{(n)}_{A})=(1+l_{IR}\delta(y-L)+l_{UV}\delta(y))m^{2}_{n}f^{(n)}_{A} \ ,
\label{diffeq}
\end{equation}
where $f_A^{(n)}(y)$ are the wavefunctions that satisfy the ortho-normality condition,
\begin{equation}
\int_{0}^{L}dy\left[1+l_{IR}\delta(y)+l_{UV}\delta(y-L)\right]f^{(n)}_{A}f^{(m)}_{A}=\delta_{nm} \ .
\label{BLKTorthonormality}
\end{equation}
The solution to Eq.\ref{diffeq} is given by,
\begin{equation}
f^{(n)}_{A}(y)=N^{(n)}_{A}e^{ky}\left[J_{1}\left( \frac{m_{n}}{k e^{-ky}}\right) +b^{(n)}_{A}Y_{1}\left( \frac{m_{n}}{k e^{-ky}} \right)\right] \ ,
\label{blktfa}
\end{equation}
where $N^{(n)}_{A}$ is the normalisation constants and $b_A^{(n)}$ the integration constant fixed by the modified boundary conditions,
\begin{eqnarray}
\partial_{y}f^{(n)}_{A}|_{y=0} &=&  - l_{UV}m_{n}^{2}f^{(n)}_{A}(0)\nonumber \ ,\\
\partial_{y}f^{(n)}_{A}|_{y=L} &=&  + e^{2kL}l_{IR}m_{n}^{2}f^{(n)}_{A}(L).
\label{EigenSolve}
\end{eqnarray}
Using the above relations, $b_{A}^{(n)}$ becomes,
\begin{eqnarray}
b_{A}^{(n)} &=& - \frac{J_{0}(\frac{m_{n}}{k})+m_{n}l_{UV}J_{1}(\frac{m_{n}}{k})}{Y_{0}(\frac{m_{n}}{k})+m_{n}l_{UV}Y_{1}(\frac{m_{n}}{k})} \ \ \text{at} \ y=0 , \nonumber \\
b_{A}^{(n)} &=& - \frac{J_{0}(\frac{m_{n}}{k}e^{kL})-m_{n}l_{IR}e^{kL}J_{1}(\frac{m_{n}}{k}e^{kL})}{Y_{0}(\frac{m_{n}}{k}e^{kL})-m_{n}l_{IR}e^{kL}Y_{1}(\frac{m_{n}}{k}e^{kL})} \ \ \text{at} \ y=0 \nonumber \\
\label{eq:gaugebcconsts}
\end{eqnarray}
where $m_{n}=x_{n}ke^{-kL}$, and $x_{n}$ are the roots of the master equation obtained by equating the two relations in Eq.\ref{eq:gaugebcconsts}. 

From Eq.\ref{diffeq} and Eq.\ref{BLKTorthonormality}, the zero mode wave-function can be derived as,
\begin{eqnarray}
f^{(0)}_{A}(y)=\frac{1}{\sqrt{L+l_{IR}+l_{UV}}} \ .
\end{eqnarray}

\begin{figure}[h!]
\vspace{-2 cm}
	\centering
     \begin{subfigure}[t!]{0.44\textwidth}
	\includegraphics[height=12 cm,width=8cm]{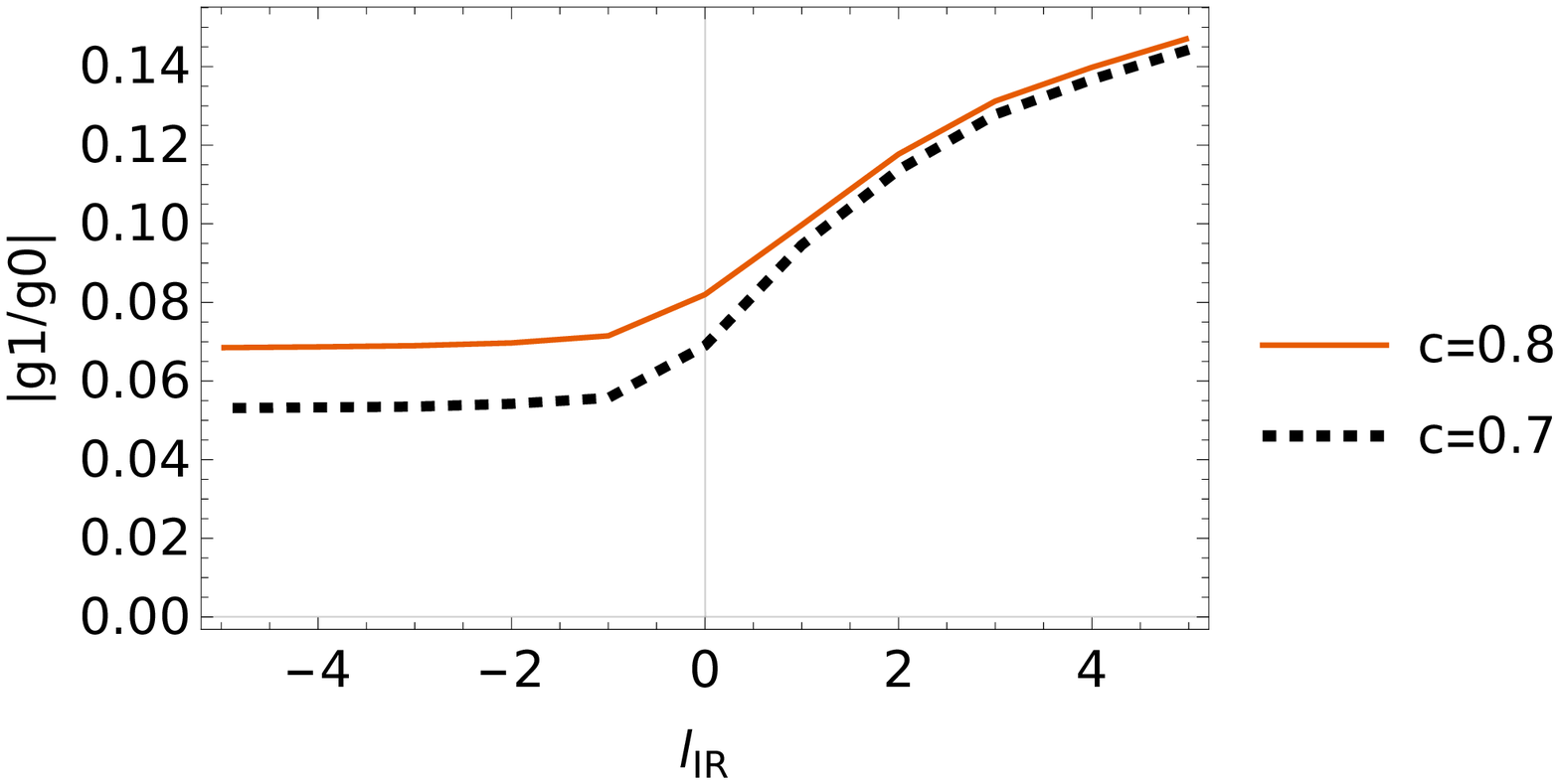}
\vspace{-4 cm}
\caption{}
\vspace{-3 cm}
     \end{subfigure}
     \begin{subfigure}[t!]{0.44\textwidth}
		\includegraphics[height=12 cm,width=8cm]{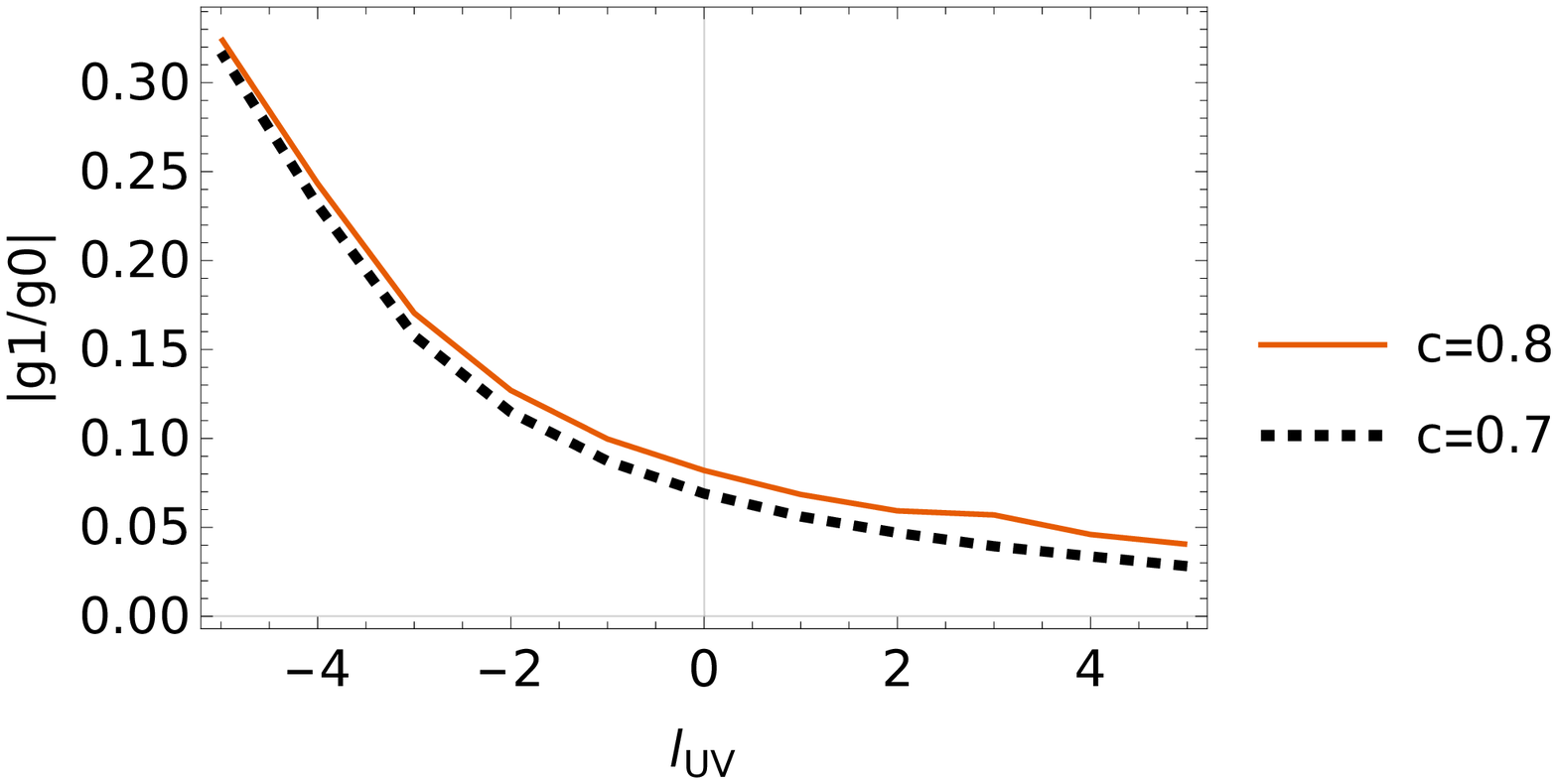}
\vspace{-4 cm}
\caption{}
     \end{subfigure}
	\caption{ (a) The coupling of KK-1 Z boson, $|g_1/g_0|$, with leptons having bulk mass parameters $c=0.8$(red solid) and $c=0.7$(black dotted) as a function of $kl_{IR}$ (assuming $kl_{UV} = 0$ ). (b) The coupling of KK-1 Z boson, $|g_1/g_0|$,with fermions having bulk mass parameters $c=0.8$(red solid) and $c=0.7$(black dotted) as a function of $kl_{UV}$ (assuming $kl_{IR} = 0$).}\label{G1G0Couplings}. 
\label{EigenValu}
\end{figure}

Since the BLKT modifies the bulk gauge field wave profile, their overlap with the lepton bilinear becomes, 
\begin{equation}
g_{L,R}^{(n)}(c_{\ell,e})=\frac{g_5}{L} \int_0^{L} dy \,e^{-3 k y} f_A^{(n)}(y) f_{L,R}^{(0)}(y,c_{\ell,e}) f_{L,R}^{(0)}(y,c_{\ell,e})\ ,
\label{coup_BLKT}
\end{equation}
where $g_5=g_0\sqrt{L+l_{IR}+l_{UV}}$, with $g_0$ denoting the coupling of $Z^{(0)}$ boson with the fermion.
This generalization of the gauge field does not affect the gauge zero-mode couplings.
On the other hand, the higher KK-mode couplings are modified significantly. Since the cLFV is mediated by the $Z_\mu^{(1)}$, to capture the effect of BLKT, it is instructive to define the quantity
\begin{equation}
\Big|\frac{\Delta g_1}{g_0}\Big| = \Big|\frac{g_{L,R}^{(1)}(c_2)-g_{L,R}^{(1)}(c_1)}{g_0}\Big| \ ,
\label{deltag1}
\end{equation}
where $g_{L,R}^{(1)}(c)$ is given in Eq.\ref{coup_BLKT}. 

For illustration, in Fig.\ref{G1G0Couplings} we display the coupling strengths of KK-1 partner of the Z boson, $\Big| \frac{g_1}{g_0}\Big|$(we denote $g_{L,R}^{(1)}(c)=g_1$), and lepton bilinears, with the bulk mass parameters $c_{1}=0.8$ and $c_{2}=0.7$, as a function of the BLKT strengths.
The values of $\Big| \frac{\Delta g_1}{g_0} \Big|$ for different BLKTs are given in Table \ref{numericalcoupling}.

\begin{table}[h]
	\renewcommand{\arraystretch}{0.9}
	\begin{tabular}{|m{0.9cm}|c|c|c|}
		    \hline
		    & & & \\[-1 ex]
	    	    BLKT &$kl_{IR}=kl_{UV}=0$&$kl_{IR}=-5$,$kl_{UV}=0$&$kl_{IR}=0$,$kl_{UV}=-5$ \\[1ex]
		\hline
		 & & & \\[-1 ex]
		$\Big| \frac{\Delta g_1}{g_0} \Big|$ & 0.013& 0.015 & 0.008 \\[1ex]
		\hline
	\end{tabular}\\[1 ex]
	\begin{tabular}{|m{0.9cm}|c|c|c|}
		    \hline
		     & & & \\[-1 ex]
	    	    BLKT & $kl_{IR}=5$, $kl_{UV}=0$&$kl_{IR}=0$, $kl_{UV}=5$&$kl_{IR}=5$, $kl_{UV}=-5$  \\[1ex]
		\hline
		 & & & \\[-1 ex]
		$\Big| \frac{\Delta g_1}{g_0} \Big|$ &0.003&0.0123&0.002 \\[1ex]
		\hline
	\end{tabular}
	\caption{$\Big| \frac{\Delta g_1}{g_0} \Big|$ values for the three cases of BLKTs computed with $c_{1}=0.8$ and $c_{2}=0.7$.}
	\label{numericalcoupling}
\end{table}

\begin{figure}[h]
\centering
     \begin{subfigure}[t!]{0.4\textwidth}
	\includegraphics[height=6.5 cm,width=8cm]{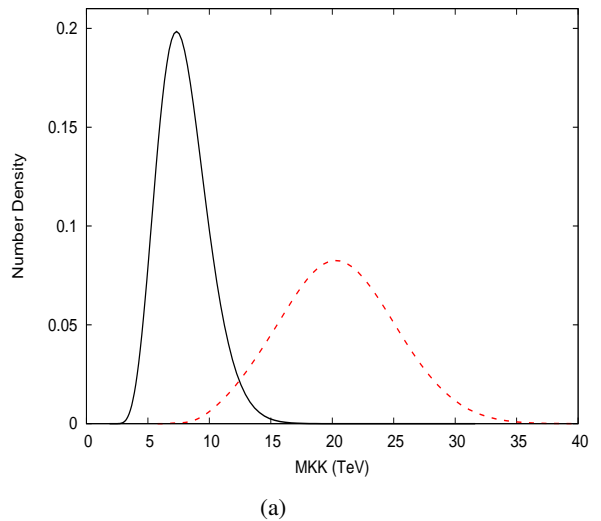}
	\caption{}
     \end{subfigure}
     \begin{subfigure}[t!]{0.44\textwidth}
		\includegraphics[height=6.5cm,width=8cm]{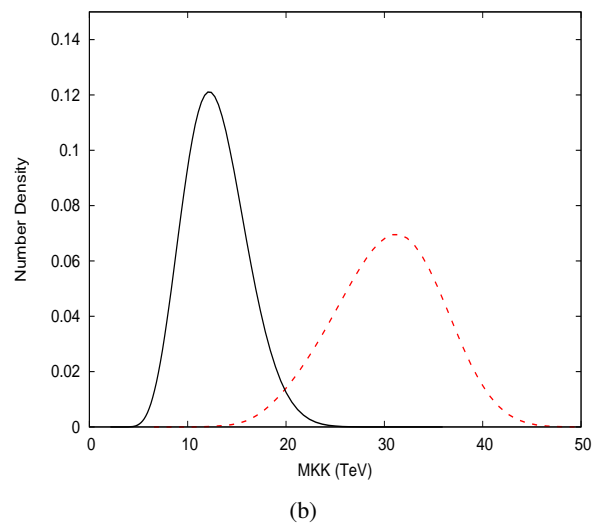}
	\caption{}
     \end{subfigure}
	\caption{ Probability distribution function satisfying (a) the experimental branching ratio for the process $\mu \rightarrow e e e$ with $kl_{IR}=5,kl_{UV}=-5$ (Black solid) $kl_{UV}=0,kl_{IR}=0$ (Red dashed) (b) the experimental $\mu Ti\to e Ti$ conversion rate $kl_{IR}=5,kl_{UV}=-5$(Black solid)$kl_{UV}=0,kl_{IR}=0$ (Red dashed)}
	\label{EigenValu}
\end{figure}

The numerical analysis is similar to the scenario without BLKT, but using the modified couplings in Eq.\ref{coup_BLKT}. The probability distribution function of simulated data, for the processes $\mu \to e e e$ and $\mu Ti \to e Ti$, is presented in Fig.\ref{EigenValu}. And for comparison, we have also shown the scenario without BLKT. These plots clearly show that the branching ratio and conversion rate constraints relax on imposing BLKT, bringing down the lower limit on the KK-1 gauge boson mass scale to $\sim 12 TeV$. Thus, making the model relevant at the upcoming hadronic collider searches. The bounds on these processes in presence of BLKT are summarised in Table \ref{Benchmark}. 

\begin{table}[h]
\centering
	\renewcommand{\arraystretch}{1.4}
	\begin{tabular}{|c|c|c|}
		    \hline
	    	  &  $kl_{IR}=kl_{UV}=0$ & $kl_{IR}= 5$,$kl_{UV}=-5$   \\
		\hline
		
	$B_{(\mu\rightarrow 3e)}$&20.8 TeV & 7.49 TeV  \\
	\hline
	$B_{\mu Ti \to e Ti}$&30.7 TeV&12.03 TeV\\
	\hline
	\end{tabular}
	\caption{Bounds on the $M_{KK}$ for the different BLKTs considered. }
	\label{Benchmark}
\end{table}


\section{Summary}
\label{sec:summary}
\begin{figure}[h]
\centering
     \begin{subfigure}[t!]{0.44\textwidth}
	\includegraphics[height=6.5cm,width=8cm]{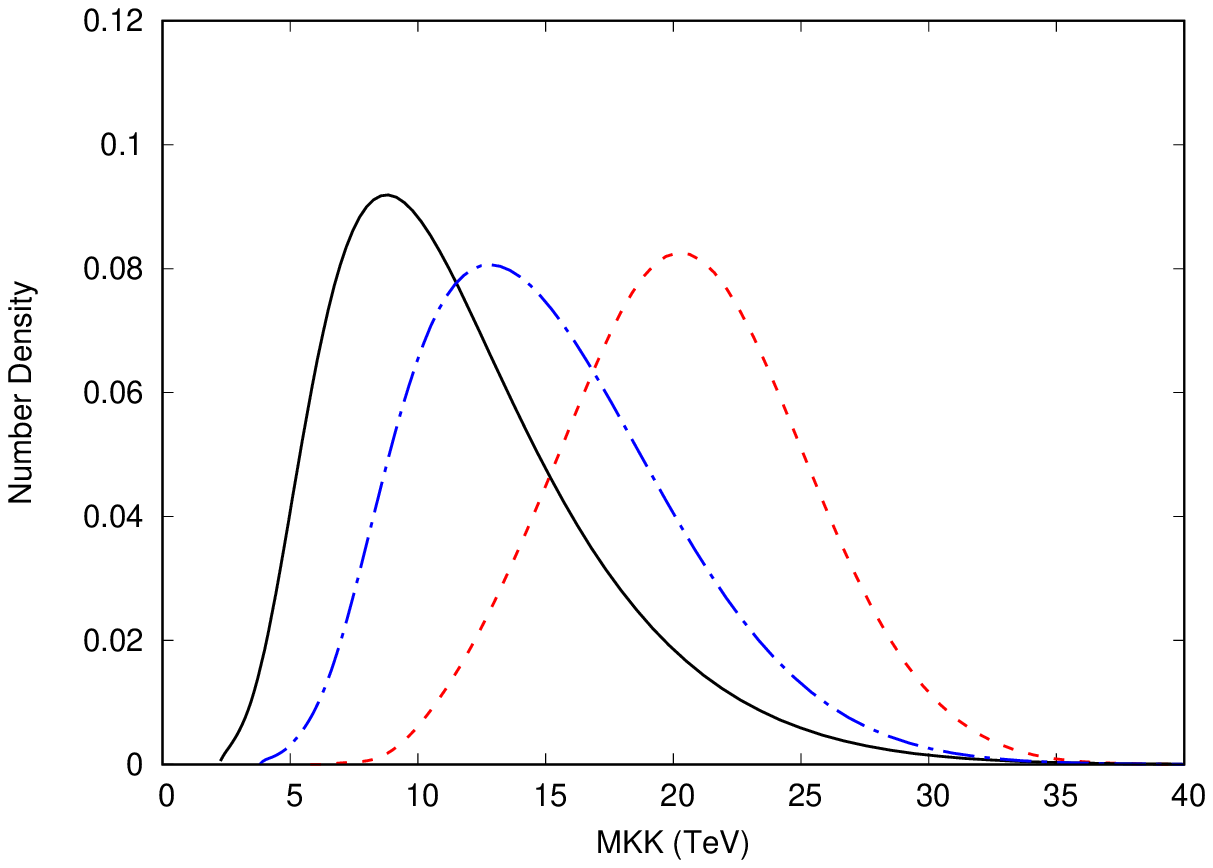}
	\caption{}
     \end{subfigure}
     \begin{subfigure}[t!]{0.44\textwidth}
		\includegraphics[height=6.5cm,width=8cm]{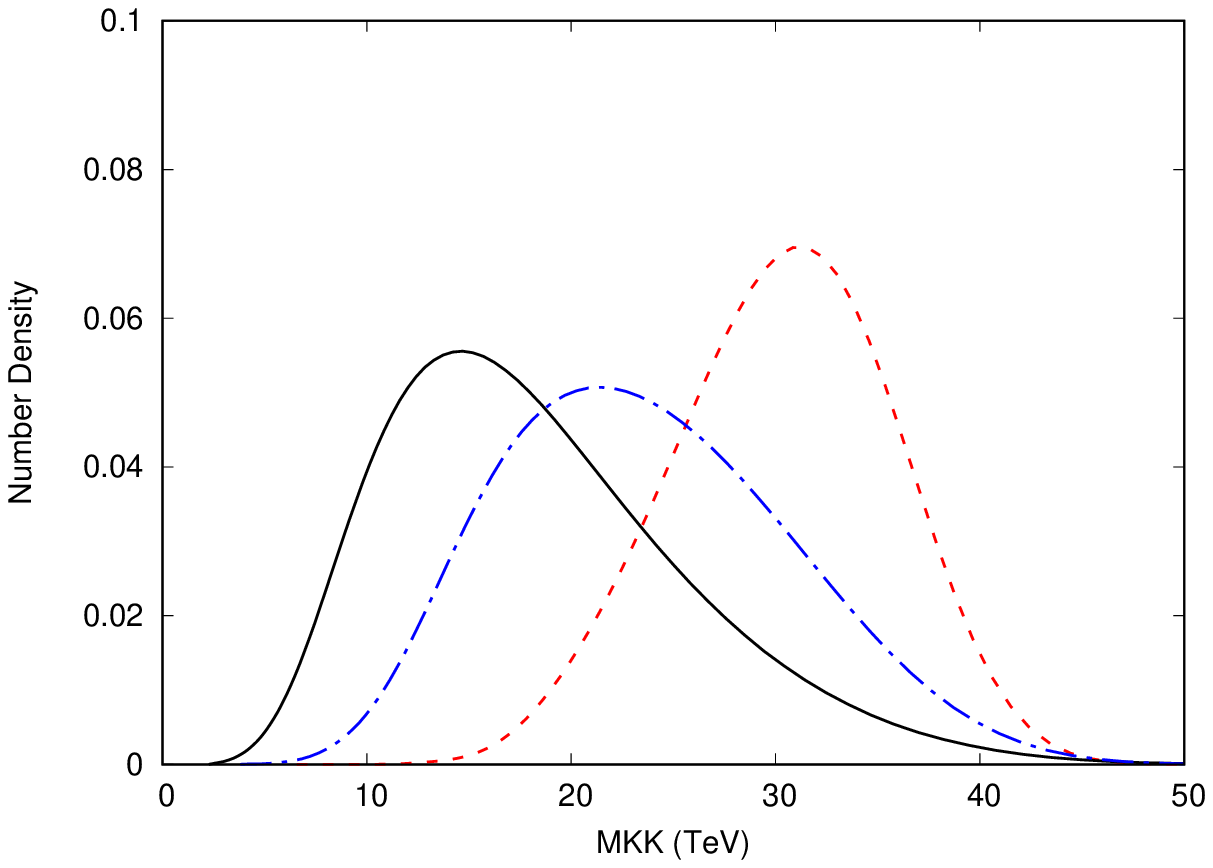}
	\caption{}
     \end{subfigure}
	\caption{ Probability distribution function satisfying (a) Constraints on the branching ratio of $\mu \rightarrow e e e$ with fat brane of size ratio 0.2 (black solid) and 0.1 (blue dot-dashed), and thin brane (red dashed) (b)Constraints on the $\mu-e$ conversion with fat brane of size ratio 0.2 (black solid) and 0.1 (blue dot-dashed), and thin brane (red dashed)}
	\label{fig:fatbrane}
\end{figure}

As an effective theory below $10^3 TeV$, the Little Randall-Sundrum model has been quite successful in relaxing the strong constraints from the electro-weak precision observables without introducing custodial symmetry. As an added advantage, it also predicts an enhanced signal at LHC\cite{DAmbrosio:2020ngh} compared to its UV complete counterpart. On the other hand, the flavour predictions turn for the worse. In a previous article by the author\cite{DAmbrosio:2020ngh}, the effect of Little RS on Kaon oscillation were discussed, where they observed that the contribution to the $\epsilon_K$ parameter was enhanced by tree level KK-1 gluon exchange diagram. The lower limit on compactification scale was computed to be $\sim 32 TeV$. On the other hand, these limits were relaxed on including the BLKT for gluons and the scale was lowered to $\sim 5 TeV$. The relaxation of the constraints were also achieved by imposing Minimal Flavour Protection, $U(3)$ flavour symmetry. This brings us to believe that the simplistic structure of gauge kinetic terms, considered so far in the literature, may not be the correct nature of the universe. Instead, we need to introduce BLKT. These generalised gauge kinetic terms are also necessary to correctly re-normalise the gauge sector~\cite{Georgi:2000wb,Arkani-Hamed:2001uol,Carena:2002dz,GrootNibbelink:2005vi,Cornell:2012qf}.

In this paper, we extend the study of flavour violation to the charged lepton sector. The anarchic Little RS model was subjected to a set of experimental constraints from the rare decays of, $\mu \rightarrow e e e $, $\tau \rightarrow e e e $, $\tau \rightarrow \mu \mu \mu $, $\tau \rightarrow \mu e e $, $\tau \rightarrow e \mu \mu $ , $\mu \rightarrow e \gamma $, and  $\mu Ti \rightarrow e Ti $. 
Unlike the case for hadrons, here, the flavour violations are mediated by the SM Z-boson through mixing with the KK-1 partners in gauge basis. We note that the Little RS suffer stronger bounds from the lepton flavour violating sector, which constrains the lower limit on the KK-1 gauge boson mass to $M_{KK} \sim 30.7 TeV$. To mitigate such large constraints and make the model viable at colliders, in this article, we proposed Brane Localised Kinetic Terms for electroweak gauge bosons. Here, we considered non-minimal kinetic terms in both UV and IR branes and found that positive values of BLKT on IR-brane and negative BLKT on UV-brane relax the bounds, effectively reducing the lower limit to $M_{KK}\geq 12 TeV$. The electro-weak sector in Little RS exhibits a very rich flavour phenomenology. These interactions, discussed in our paper, can also contribute significantly to processes involving Lepton Universality violation.

Before concluding, we note that that the flavour violations, discussed here, are generated due to the difference in the localisating wave profiles of fermion zero modes at the UV-brane. If the wave profiles were degenerate, the KK modes of the $Z$ boson would have coupled democratically to all the leptons. It is interesting to investigate whether these large corrections can be mitigated by relaxing our assumption on the unnatural thinness and rigidity of the brane. Though beyond the scope of this paper, if we consider fat branes, we can show that the lower limit on $M_{KK}$ softens significantly. For demonstration, we choose two scenarios with the ratio of brane width to compactification radius 0.1 and 0.2, and we have plotted the Probability Distribution Function satisfying the constraints on the branching ratio of $\mu \rightarrow e e e$ and $\mu \to e$ conversion in Fig.\ref{fig:fatbrane}. For ratio 0.2, it can be seen that the constraints from $\mu \rightarrow e e e$ limits $M_{KK} \gtrsim 8 TeV$, while $\mu Ti \rightarrow e Ti$ limits $M_{KK} \gtrsim 12 TeV$. Moreover, for flavour violations arising on the brane, the pseudo-Nambu-Goldstone bosons of the spontaneously broken translational symmetry (branons) are understood \cite{Bando:1999di,Arun:2019ogf} to suppress the coupling of gauge boson KK-modes with the fermion bilinear. Hence, a major part of the flavour violation appears outside the brane. This interesting feature disappears as the brane gets thinner and more rigid.

\section*{Acknowledgements}
M.T.A. acknowledges financial support of DST through INSPIRE Faculty grant [DST/INSPIRE/04/2019/002507].
\section{Appendix}
\subsection{Mass Matrix of Gauge Fields}
\label{sec:gaugemass}
After spontaneous symmetry breaking of the Higgs, the mass terms from the Lagrangian given in Eq.\ref{Lgauge} and in Eq.\ref{Lhiggs} becomes,
\begin{equation}
\begin{split}
&\mathcal{L}_{m}=\sum_{n}m^{(n)2}_{w}W^{+(n)}_{\mu}W^{-\mu(n)}+m_{A}^{(n)2}A^{(n)}_{\mu}A^{\mu(n)}\\&+m_{z}^{(n)2}Z^{(n)}_{\mu}Z^{\mu(n)}+\frac{v^{2}}{2}  \int dy \delta(y-L)  g_{5}^{2} e^{-2ky}\times\\&\left(\sum_{m,n} W^{+(m)}_{\nu}W^{-\mu(n)}f_{w}^{(m)}(y)f_{w}^{(n)}(y))\right)+\frac{v^{2}}{2}(g_{5}^{2}+g_{5}'^{2})\times\\&\int e^{-2ky}\sum_{m,n}Z^{(m)}_{\mu}Z^{\mu(n)}f_{z}^{(m)}(y)f_{z}^{(n)}(y)\delta(y-L) dy,
\end{split}
\label{masslag}
\end{equation}
where $m^{(n)}_{w}$, $m^{(n)}_{A}$ and $m^{(n)}_{z}$ is the nth KK masses of W-boson, photon, Z-boson respectively.\\
Mass matrix of these gauge fields computed from the Lagrangian Eq.\ref{masslag}:\\
The mass term of $A_{\mu}$ is,
\begin{equation}
\label{massmatrixA}
\begin{bmatrix}A_{\mu}^{(0)}&A_{\mu}^{(1)}\end{bmatrix}\begin{bmatrix}
0&0\\0&m^{(1)2}_{A}\end{bmatrix}\begin{bmatrix}A_{\mu}^{(0)}&A_{\mu}^{(1)}.
\end{bmatrix}
\end{equation}
Zero-mode of the photon does not couple with the Higgs and hence it is massless.  \\
\\
The mass term of $W^{\pm}_{\mu}$ is,  \\
\begin{equation}
\begin{split}
 &M_{w}= m^{(0)2}_{w}W^{+(0)}_{\mu}W^{-\mu(0)} +m^{(1)2}_{w}W^{+\mu(1)}_{\mu}W^{-\mu(1)}+\frac{v^{2}}{2} \times \\&\int g_{5}^{2} e^{-2ky}\sum_{m,n} W^{+(m)}_{\mu}W^{-\mu(n)}f_{w}^{(m)}(y)f_{w}^{(n)}(y)\delta(y-L)dy.
\end{split}
\end{equation}
This can be  represented by the matrix form,
\begin{equation}
\label{massmatrixw}
M_{W}=\begin{bmatrix}W_{\mu}^{+(0)}&W_{\mu}^{+(1)}\end{bmatrix}\begin{bmatrix}a_{00}&a_{01}\\a_{10}&a_{11}+m^{(1)2}_{w}\end{bmatrix}
\begin{bmatrix}W_{\mu}^{-(0)}\\W_{\mu}^{-(1)},
\end{bmatrix}
\end{equation}
where $a_{m,n}=g_{5}^{2}\frac{v^{2}}{2}e^{-2kL}f_{w}^{(m)}(L)f_{w}^{(n)}(L)$.\\
\\
Similiarly mass term of the Z boson is,
\begin{equation}
\begin{split}
&M_{Z}= m^{(0)2}_{z}Z^{\mu(0)}_{\mu}Z^{\mu(0)} +m^{(1)2}_{z}Z^{(1)}_{\mu}Z^{\mu(1)}+\frac{v^{2}}{2}(g_{5}^{2}+g_{5}'^{2})\times \\&\int e^{-2ky}\sum_{m,n}Z^{(m)}_{\mu}Z^{\mu(n)}f_{z}{(m)}(y)f_{z}^{(n)}(y)\delta(y-L) dy.
\end{split}
\end{equation}
\begin{equation}
\label{massmatrixz}
M_{Z}=\begin{bmatrix}Z_{\mu}^{(0)}&Z_{\mu}^{(1)}\end{bmatrix}\begin{bmatrix}b_{00}&b_{01}\\b_{10}&b_{11}+m^{(1)2}_{z}\end{bmatrix}\begin{bmatrix}Z_{\mu}^{(0)}\\Z_{\mu}^{(1)}
\end{bmatrix},
\end{equation}
where $b_{m,n}=g_{5}^{2}\frac{v^{2}}{2}e^{-2kL}f_{z}^{(m)}(L)f_{z}^{(n)}(L)$ and $Z^{\mu (0)}$, $Z^{\mu (1)}$ are in the gauge basis of the Z boson. This is a feature of IR-brane localised electro-weak symmetry breaking.
\subsection{Couplings and $l_i \to 3 l_j$ branching ratio}
\label{subsec:coupbr}
On going to mass basis, the flavour-basis coupling matrices $C^{F}_{L,R} = g_{L,R} \, {\rm diag}(\alpha_e,\alpha_{\mu},\alpha_{\tau})$ gets rotated to 
$C_{L,R}=U_{L,R} \,C^{F}_{L,R} \,U^{\dag}_{L,R}$. Since $\alpha_e \neq \alpha_\mu \neq \alpha_\tau$, in the rotated basis $C_{L,R}$ generate off-diagonal entries that lead to flavour violation.
Using the unitarity of $U_{L,R}$, we get,
\begin{equation*}
    g_{L,R}^{(1)\mu e}= g_{L,R}(U_{1,2}^{L,R}U_{2,2}^{L,R*}(\alpha_{\mu}-\alpha_e)+U_{1,3}^{L,R}U_{2,3}^{L,R*}(\alpha_{\tau}-\alpha_e) ) \ ,
\end{equation*}
\begin{equation*}
    g_{L,R}^{(1)\tau\mu}= g_{L,R}(U_{2,1}^{L,R}U_{3,1}^{L,R*}(\alpha_e-\alpha_{\mu})+U_{2,3}^{L,R}U_{3,3}^{L,R*}(\alpha_{\tau}-\alpha_e) ) \  ,
\end{equation*}
\begin{equation}
    g_{L,R}^{(1)\tau e}= g_{L,R}(U_{1,2}^{L,R}U_{3,2}^{L,R*}(\alpha_{\mu}-\alpha_e)+U_{1,3}^{L,R}U_{3,3}^{L,R*}(\alpha_{\tau}-\alpha_e) ) \ .
\end{equation}
where $g_{L,R}$ are the usual SM couplings. Using Eq.\ref{gaugemix}, the couplings to $Z_0$ are obtained via multiplication by $f\frac{m_Z^2}{M_{KK}^{2}}$, that is
\begin{equation}
g^{l_i l_j}_{L,R} =\frac{-fm_{z}^{2}}{M_{KK}^{2}} g^{(1)l_i l_j}_{L,R} \ ,
\end{equation}
With the above equations, it is now relatively simple to derive the Effective Lagrangian is given in Eq.\ref{ngeq}. The Wilson coefficients $g_{3,4,5,6}$ given in the Effective Lagrangian arise from the processes with the exchange of $Z_{(0)}$ and $Z_{(1)}$ bosons. These can be derived as,
\begin{eqnarray}
g^{l_{i}l_j}_3 &=& 2g_R \left[f-\alpha_j\right]\frac{m_Z^2}{M_{KK}^{2}} g_{R}^{(1) l_{i}l_j}, \nonumber \\
g^{l_{i}l_j}_4 &=& 2g_L \left[f-\alpha_j\right]\frac{m_Z^2}{M_{KK}^{2}} g_{L}^{(1) l_{i}l_j}, \nonumber \\
g^{l_{i}l_j}_5 &=& 2g_L \left[f-\alpha_j\right]\frac{m_Z^2}{M_{KK}^{2}} g_{R}^{(1) l_{i}l_j}, \nonumber \\
g^{l_{i}l_j}_6 &=& 2g_R \left[f-\alpha_j\right]\frac{m_Z^2}{M_{KK}^{2}} g_{L}^{(1) l_{i}l_j},
\label{fvcoups2}
\end{eqnarray}
where $i$ and $j$ are $(e,\mu,\tau)$.
The first term in the above equation are computed from the $Z_0$ exchange, while the second is from $Z_1$ exchange. \\
The relevant branching 
fractions for the process $\l_{i} \rightarrow 3\l_j$ now become~\cite{Chang:2005ag, Agashe:2006iy}, 

\begin{equation}
BR(\l_i \rightarrow 3l_j) = 2\left(|g^{\l_i \l_j}_3|^2+|g^{\l_i \l_j}_4|^2\right)+|g^{\l_i \l_j}_5|^2+|g^{\l_i \l_j}_6|^2 \ .
\label{bfracs1}
\end{equation}
\subsection{Numerical Analysis}
\label{subsec:appendix}
Here we present the numerical analysis of the Little RS parameter space, to determine how accurately the Little RS geometric origin of flavour can be tested in current and future experiments. We will discuss the full parameter scan of the bulk mass parameters, namely `c' values, that fit the lepton mass within the experimental error. We have assumed anarchic Yukawa in the lepton sector. We choose the basis in which the left-handed mixing matrix $U_{L}=I$ and $U_{R}$, the right-handed mixing matrix, contains the mixing elements. This means that the flavour violation are generated in the right-handed sector and hence the model is independent from fitting the $U_{PMNS}$ matrix, which should happen once neutrino phenomenology is modelled in Little RS.

The $3\times3$ complex matrices of 5-dimensional Yukawa couplings $Y_e$ contains 9 real and 9 complex elements. Since we have assumed left-handed sector to be aligned with SM, the 5-dimensional Yukawa should also be assumed to have some symmetry. Which reduces the total free parameters to 6 real and 3 complex phases. For simplicity, we also choose the basis in which the bulk mass parameters are diagonal and real. 

We restrict the 5-dimensional Yukawa couplings to the range $0.1\leq |Y_{i,j}|\leq 3$ so that the values do not introduce unnatural hierarchies and remain below the perturbative limit in the model. Ignoring the mixing of fermionic KK-modes, for the time being, we can write,
\begin{equation}
    \zeta =U_L^{\dagger}Y_{4D}U_R \ ,
\end{equation}
where, $Y_{4D}$ are the 4D Yukawa coupling defined in Eq.\ref{4dyuk} and $\zeta$ is defined as
\begin{equation}
    \zeta=\frac{\sqrt{2}}{v} diag(m_e,m_{\mu},m_{\tau}).
\end{equation}
Using this, we compute the 5-dimensional Yukawa coupling  as
\begin{equation}
    (Y_{5D})_{i,j}=f^{(0)-1}(c_{\ell i})(U_L\zeta U_R^{\dagger})f^{(0)-1}(c_{ej}) \ .
\end{equation}
Where $U_L$ and $U_R$ are the lepton mixing matrices. 
\begin{table}[h]
\centering
	\renewcommand{\arraystretch}{1.4}
	\begin{tabular}{|c|c|c|c|c|c|}
		    \hline
	    	$c_{\ell 1}$  &$c_{\ell 2}$  &$c_{\ell 3}$&$c_{e1}$&$c_{e2}$&$c_{e3}$  \\
		\hline
		
	2.15-2.25&1.1-1.2 & 1-1.05&0.8-0.95&0.65-0.8&0.55-0.70\\
	\hline
	\end{tabular}
	\caption{The of bulk mass parameter $`c'$ used in our scan.}
	\label{cvalu}
\end{table}
We have run the scan over $10^6$ iterations and collected 5000 points in the Little RS parametric space satisfying the anarchic Yukawa conditions.
The range of $c$ values satisfying the above condition given in Table\ref{cvalu}
Using these points, we calculated the bounds on $M_{KK}$ for all decays mentioned in sec.\ref{sec:trileptondecay}.

\bibliographystyle{ieeetr}
\bibliography{Notes.bib}
\end{document}